\documentclass[12pt]{article}
\usepackage{graphicx}

\begin{document}

\title{Public debates driven by incomplete scientific data: the cases of evolution theory, global warming and H1N1 pandemic influenza}
\author{Serge Galam\thanks{serge.galam@polytechnique.edu}\\
Centre de Recherche en \'Epist\'emologie Appliqu\'ee (CREA),\\
\'Ecole Polytechnique and CNRS, \\Boulevard Victor, 32,
F-75015 Paris, France}

\date{}
\maketitle

%%%%%%%%%%%%%%
\begin{abstract}

Public debates driven by incomplete scientific data where nobody can claim absolute certainty, due to current state of scientific knowledge, are studied. The cases of evolution theory, global warming and H1N1 pandemic influenza are investigated. The first two are of controversial impact while the third is more neutral and resolved. To adopt a cautious balanced attitude based on clear but inconclusive data appears to be a lose-out strategy. In contrast overstating arguments with wrong claims which cannot be scientifically refuted appear to be necessary but not sufficient to eventually win a public debate. The underlying key mechanism of these puzzling and unfortunate conclusions are identified using the Galam sequential probabilistic model of opinion dynamics  \cite{mino, hetero, inflexible}. It reveals that the existence of inflexible agents and their respective proportions are the instrumental parameters to determine the faith of incomplete scientific data public debates. Acting on one's own inflexible proportion modifies the topology of the flow diagram, which in turn can make irrelevant initial supports. On the contrary focusing on open-minded agents may be useless given some topologies. When the evidence is not as strong as claimed, the inflexibles rather than the data are found to drive the opinion of the population. The results shed a new but disturbing light on Designing adequate strategies to win a public debate.

\end{abstract}
 
%\pacs{02.50.Ey, 05.40.-a, 89.65.-s, 89.75.-k, 05.00.00 Statistical physics, thermodynamics, and nonlinear dynamical systems; 89.20.-a Interdisciplinary applications of physics; 89.75.-k Complex systems}

\maketitle

%%%%%%%%%%%%%%
\newpage

%%%%%%%%%%
%%%%%%%%%%%%%
\section{Setting the problem}

Public opinion is today a key ingredient  in modern societies policy making. To discover the laws and eventual biases which govern its forming is thus a major challenge to avoid  political choices founded on distorted expressions from what is the overall majority of individual opinions  \cite{fortunato-review, deffuant, sznajd, mino, slanina, redner-2, frank-voter, espagnol,  schneider, lambiotte-ausloos, ausloos-religion, lambiotte-ausloos-holyst, pierluigi, mariage, martins1, martins2, galam-review}. Any progress in the understanding of opinion dynamics could have drastic effects on the way  sensitive issues are tackled. 

Among them stand the public debates initiated from ``science" in presence of incomplete scientific data. They are driven by the necessity to fill up the missing parts to come up with a complete view of the issue from which rationally motivated political decisions can be made. It is therefore of central importance to determine if what is accepted as the complete theory corresponds to the real scientific status of the issue. Evolution theory, global warming and  H1N1 pandemic influenza debates are emblematic of this class of public issues. Accordingly we investigate the conditions in which those debates were set with an emphasis on the scientific facts, the scientist positions, the associated general beliefs, the political stakes and the resulting public opinions. 

First, we examine the two cases of Evolution theory and global warming, which both have controversial impact. Similarities and differences are enumerated as well as some apparent contradictions between the respective outcomes. A throughout analysis reveals that to adopt a cautious balanced attitude based on clear but inconclusive scientific facts is a lose-out strategy. In contrast, overstating arguments and asserting wrong statements which cannot be scientifically refuted, is found to be necessary but not sufficient to eventually win a public debate. 

These findings are then confronted to the more neutral case of the H1N1 pandemic influenza, which has neither emotional nor religious character. Moreover, it has the advantage of being over with available data on the debate final outcome together with both  opposed side retrospective stances.

On this basis, to identify the underlying key mechanism, which produces these puzzling but unfortunate conclusions, we use the Galam sequential probabilistic model of opinion dynamics  \cite{mino, hetero, inflexible}. It is shown that the existence of inflexible agents, i.e., agents who never change their opinion, is the instrumental parameter to determine the faith of a public debate based on incomplete data. Depending on their respective densities the associated opinion forming obeys either a non threshold or a threshold dynamics. Initial supports may thus turn to be irrelevant. 

The results shed a new but disturbing  light on Designing adequate strategies to eventually win public debates. To produce inflexibles in one's own side is thus critical to win a public argument  whatever the rigor cost and the associated epistemological paradoxes.
At odds, to focus on convincing open-minded agents is useless. In summary, when the scientific evidence is not as strong as claimed, the inflexibles rather than the data are found to drive the collective opinion of the population. Consequences on Designing adequate strategies to win a public debate are discussed.

\section{Revisiting evolution and global warming issues}
\label{visit}

The issues of global warming and evolution theory have triggered intense public debates to enforce eventual political decisions in order to modify fundamental setting in the society organization. Both issues are concerned with the understanding of global phenomena whose characteristic timescales are extremely long. However, only scarce and incomplete data are available. It was thus necessary to build ad hoc theories to produce a coherent explanation of the whole respective fields, embodying isolated facts, fuzzy measures and partial knowledge.  To bridge the missing links, complicated models have been elaborated under the constraint of reproducing all past known data \cite{models}. Ultimately, large scale simulations were performed on big computers to make predictions about possible evolution in the future as well as Designing new research programs with guidelines for research funding. 

Nevertheless, these theories are suffering the basic fact that up to date no global science has been established in any field of research. The recent effort to develop the so called science of complex systems aims to remedy these deficiencies \cite{complex}. But yet without any major breakthrough so far. Present ``sciences" are restricted to local domains and can produce scientific evidence only for isolated phenomena. Along these difficulties, it is worth to stress that the major limitation of both evolution and climatology theories is their incapacity to make predictions, which could be refuted by setting an experiment. This is due to the huge timescales involved in each case, of the order of millions of years for evolution and hundreds of years for the climate. It thus prevents these theories to reach the status of hard science like physics and chemistry although they do use scientific methods. They could possibly reach this status in a few hundreds of years after a series of successful predictions, which would allow to evaluate a degree of reliability.

Current data being incomplete, a large amount of speculation is incorporated to compensate the missing parts, and numerous assumptions are made, which cannot be checked. Therefore any result obtained from the numerical simulations should be taken with extreme caution in particular if political decisions are to be implemented.
 
Accordingly, scientists working on these fields, while dealing with some specific questions, often use the expression ``we believe that...", which states a clear difference with ``it is proved that..." \cite{belief1}. Unfortunately the same caution often does not hold while defining the key mechanism at work behind the phenomenon \cite{climate1}. Natural selection is asserted to be the operating criterion for evolution and anthropogenic carbon dioxide  is designated as the cause of recent global warming. Both assertions being stated as if they were  scientifically proved. Up to the present moment they constitute the most convenient key to yield a coherent picture of the respective field but this does not constitute a scientific proof of their validity. 

Such overestimates of the scientific validity of a theoretical hypothesis could be without any solid consequence like for instance with the big bang theory. To believe that the big bang hypothesis is proven scientifically does not imply anything neither in the organization of our societies nor in our personal lives. Indeed, no one is drawing social consequences  from the supposed existence of the big bang. No ``political" complain was neither made so far against its assumption.

In contrast evolution and climate issues have direct political implications making their status and validity an instrumental key to set up decisions on very crucial and sensitive political arbitrages. While the climate trend deals with the planet future and thus with the human survival on the planet, evolution touches to the sensitive question of the existence of God. Accordingly confusion about what is believed to be true and what has been proven to be true reveal itself to be problematic with rather controversial impact. It could drive political decisions followed with social catastrophes \cite{plon}. 

The case of H1N1 pandemic influenza is treated separately in Section \ref{A} since the public debate is over and of a different nature. 

\subsection{Evolution theory}

Evolution theory being taught in high schools it could be perceived as putting at stake the genesis description of the world creation. Religious people could thus feel interfered in their beliefs by the teaching of Darwin theory, although creationism has been dismissed by science and prohibited from being taught as science by the 1987 US Supreme Court ruling  \cite{highcourt1}. However creationists did not give up their refusal of the natural selection and recently they launched a massive campaign to relativize Darwin theory with a challenging theory called the Intelligent Design \cite{intelligent1, intelligent2}. It is indeed a creationist view of evolution dressed under a scientific set up  \cite{intelligent3}. 

They succeeded in initiating a vast public debate with the requirement to have  Intelligent Design taught in public schools on the same footing as Darwin theory. After a fierce battle, a federal judge for the U.S. District Court has decided in 2005 against the request dismissing the scientific character of the Intelligent Design \cite{highcourt2}. During the impassioned debate, advocates of Darwin theory opposed the  Intelligent Design view asserting Darwin theory is scientifically proved \cite{darwin}. They did win both the legal issue and the public debate with the support of a majority of the public, but yet with a substantial part still abiding along the Intelligent Design frame  \cite{ausloos}.

\subsection{Global warming}

The global warming situation differs for climatology, which up to thirty years ago was not among the first priorities neither in science nor in society. It has been the scary feeling due to an increase in natural catastrophes (this increase is difficult to evaluate precisely), combined with a so called anomalous increase of the earth global temperature from 1978 till 1998 followed by a plateau and a slightly decrease in the last years, which has propelled climate evolution at the top of world government agendas \cite{climate1}. 

Contrary to evolution theory it has been the current ground reality which has prompted a public debate to determine both the cause of the global warming and the means to curb it. Rather quickly a new paradigm has been established to embrace the phenomenon. The increasing production of carbon dioxide from man activities has been claimed to be identified as the cause of global warming and its drastic and immediate reduction set as the clear solution  \cite{climate1}. 

Only a few skeptics dispersed over the world have resisted this explanation but failed in gaining a substantial public support  \cite{climate-polls}. The alarmists have won the public debate asserting their claim is scientifically proved. Soon they may well have world governments to abide along their requests with a profound rearrangement of our way of life. For an in depth analysis of the social aspect of the global warming phenomenon  see \cite{plon}.

\subsection{Similarities between the two issues}

\begin{itemize}
\item
The core and the majority of involved scientists in each field and more generally scientists from all fields support the claim that the identified major mechanisms at work, respectively Darwin natural selection and anthropogenic carbon dioxide, have been scientifically established.

\item
Both communities have won the public debate against their opponents, respectively  Intelligent Design advocates and human-caused global warming skeptics.

\item
Both groups convinced policy makers to respectively forbid the teaching of  Intelligent Design and to oblige a drastic decrease in carbon dioxide emissions to curb global warming.
\end{itemize}

\subsection{Differences between the two issues}

\begin{itemize}
\item
A substantial part of the public still support  Intelligent Design while no substantial support oppose the global warming human responsibility. Although this last fact has been changing in the US recently, due in part to last years slight global temperature decrease \cite{climate3}.

\item Intelligent Design supporters took the initiative to launch a public debate and scientists had to oppose it. In contrast, climatologists took the initiative to alert the public on global warming with only a few scientists to oppose them.

\item
Up to 2000 most people did not consider the climate is an issue and did not look at carbon dioxide emissions as causing global warming. The situation has heated up with the 2007 publication of the IPCC  \cite{climate1}.
\item
It is the opponents to Darwin theory who were asking changes while its advocates have no peculiar demand beside to reject their request as unfounded scientifically. In contrary it is the proponents of global warming man responsibility who are demanding drastic changes in our way of life with a substantial cut in our carbon dioxide emissions while the skeptics are opposing the claimed man responsibility.

\item
Supporters of  Intelligent Design are convinced that the true explanation of evolution resides in a superior entity while climate skeptics are only pointing to the absence of a scientific proof of the human responsibility in the global warming.

\end{itemize}

%%%%%%%%%%

\section{The paradoxical questions}

In the case of evolution theory, the quasi totally of the scientific community was united to oppose the tentative to incorporate  Intelligent Design as an alternative to Darwin theory in the educational curriculum \cite{intelligent1, intelligent2}. It succeeded in dismissing the alleged scientific character of  Intelligent Design and thus did convince institutions and a majority of the public to reject the call for changing school programs not mixing science and religion. However they failed to convince a substantial part of the public for which Intelligent Design is as scientific as Darwin theory. Several puzzling questions arise :

\begin{itemize}
\item
Why in order to refute scientifically  Intelligent Design did scientists sometimes overdo the scientific validity of Darwin theory? 
\item
Why did scientists fail to convince such a large proportion of the population that Intelligent Design is not science?
\end{itemize}

The rigorous attitude would have been to state that Darwin theory is the best frame up to date to explain evolution in conformity with available data, but one cannot claim it has been scientifically proven. Such a tempered position is not contradictory with proving  Intelligent Design is not science compatible.

With respect to global warming a large part of the scientific community has supported the climatologists launching a public alert about the immediate emergency of cutting down carbon dioxide emissions to curb a coming climate driven apocalypse. The goal was to convince the public and the authorities to enforce a drastic change in our way of life. It was not like for evolution theory to oppose a hostile group holding a different view on the climate. 

And here too the claims were and are overdone, in particular about man responsibility, which is asserted by many to be scientifically proven. However the scientific ground on which this claim relies is far much thiner than for the natural selection of Darwin theory. Yet most scientists were asserting that we are in an anomalous global warming trend and man is responsible for it. They end up successful in having a large majority of the public opinion to back their view and in putting the world policy makers under a tremendous pressure to act. 

In reaction to the excess and exaggeration of the alarmist discourse an increasing number of scientists did stand to denounce such a non scientific practice. They dismissed the fears and refuted the man responsibility claim as not scientifically proved. However they did not claim to have an alternative explanation. Moreover they could not prove man is not responsible since it is impossible to prove the non-existence of something which does not exist. Skeptics failed in spreading their doubts staying confined to a tiny minority against a huge majority holding as true the man responsibility. They also failed in convincing policy makers not to get engaged in a huge program to cut carbon dioxide emissions. The questions here are:

\begin{itemize}
\item
Why did the alarmists succeed in getting the majority of public opinion to align along their unproved claim ? Even if the support has decreased recently \cite{climate3} due to the disclosure of a series of mistakes in the 2007 IPPCC report.
\item
Why they made their overstatements so dramatic while there exists no substantial opposition? 
\item
Why the tptics who adopt a rigorous scientific position without advocating an alternative claim failed to crystallize at least some part of the public opinion? 
\item
Why despite the skeptic failure the alarmist majority has been very adamant in slamming the skeptic behavior and banning them from both scientific institutions and the media?
\end{itemize}

%%%%%%%%%%

\section{The circumvented case of H1N1 pandemic influenza}
\label{A}

Above two cases have controversial impact. The legal question of evolution theory has been settled with the 2005 U.S. District Court ruling and the public debate is closed so far. But the intelligent Design people keep on their view and will certainly try to relaunch a debate once the conditions will permit. In contrast the global warming issue is still under active debate and not much has been settled yet. 

On this basis, it is fruitful to confront our findings to the H1N1 pandemic influenza case, which is more neutral with neither emotional nor religious aspects. In addition the public debate is over \cite{h1}. The case is circumvented and completed with all data now available. However there exists an essential difference with above two issues. Those are  concerned with an opinion which in principle can be reversible. In contrast, the vaccination act is irreversible. Accordingly the model can provide an explanation about the readiness among the population to get vaccinated during the vaccination campaign. Indeed many of the vaccination centers were used below their capacities to finally get closed at the beginning of 2010.

To be more precise we consider the case of France where the government had ordered more than ninety millions of individual doses of vaccine, with a result of only around six millions people vaccinated \cite{h2}. The issue was for each person to decide on wether or not to get vaccinated. Moreover it was free of charge.  Accordingly it can appear as a paradox to have such a low rate of vaccination \cite{h3}. 

Without tracing back the whole history of the H1N1 pandemic influenza we can just recall that the World Health Organization (WHO) has declared in June 2009 a pandemic situation and about 15000 people died from it \cite{h4}. However, there has been so much controversy over whether or not to get the H1N1 flu vaccine. The paradox here being that a series of people opposed adamantly the vaccination process although there was no clear danger to it, non withstanding the fact that its benefit was not proved neither.

On the other hand, people who advocated the vaccination only claimed it was a good thing to do but without stating people must do it. They only strongly recommend vaccination, even often setting an example by getting vaccinated themselves but without pretending it is a must. This subtle positioning stems from the fact that some dangerous side effect, although very rare, were not totally precluded. To order an action whose consequence is not $100\%$ sure may turn to some unwanted dangerous act making people cautious. 

In contrast people who were asserting the H1N1 pandemic influenza is not more dangerous than usual seasonal influenza virus and the vaccine not being a solid defense, did not state not to get vaccinated always mentioning it is an individual choice to be decided by each person. They casted a doubt on the prospect. It is therefore not symmetrical to dismiss the immune character of a vaccine than to order someone to get vaccinated. In particular with respect to a few millions of cases for which it is certain that a few cases could turned with bad secondary effect opening the way for possible prosecution of the people in charge.

There was much more conviction on the refusal side than on the  vaccination side and the net result among the public has been an overwhelmed majority who did not get vaccinated putting the french government with the burden of getting rid of tens of millions of vaccines. This brief overview of the case is coherent with our general findings with respect to both evolution and global warming issues. 

It is also of importance to focus on the fact that once the danger was over, a great deal of criticism has been directed against policy makers as well as members of the WHO putting light on their professional links with pharmacologic industry. Several investigations have been set to check the matter \cite{h5}.
 
 %%%%%%%%%%%

\section{Setting the model to hint at some understanding}

To tackle the problem via the prism of opinion dynamics might be fruitful to provide above  questions with a new light. Two kinds of agents seem to be involved, the ones who are claiming a truth about the issue and the ones who have no direct access to the frame of the truth and thus want to make up their mind via discussions with others. In the first category stand scientists convinced Darwin theory or global warming man responsibility has been proven scientifically as well as believers who take literally the Genesis description of the world creation. All others belong to the second category including the skeptical scientists who discard man responsibility for climate change. With respect to the H1N1 pandemic influenza scientists were present on both sides with a net majority on the refusal side.

Denoting inflexibles the first agents and floaters the others leads to evoke the Galam sequential probabilistic model of opinion dynamics \cite{mino, hetero, inflexible}. It considers a mixture of heterogeneous agents with inflexibles and floaters. Inflexibles discuss with other agents but in order to convince them of their own truth. They can eventually convince a floater to shift its opinion but they will preserve their own opinion no matter they are successful or not in convincing others. In contrast, floaters  do have an opinion about the issue but are inclined to shift opinion if eventually given sufficient convincing arguments. Using a one-person-one-argument principle, the model implements an opinion shift via small group discussions monitored by local majority rules.

\subsection{The bare local majority model}

The model consists of a group of $N$ agents undergoing a public debate. Each agent holds either one opinion A or B. Before the public debate is turned on agents did make a choice individually according to their information and belief. Initial global proportions of both opinions are respectively $p_t$ and $(1- p_t)$. They can be evaluated using polls.

The opinion dynamics is then driven by a series of repeated cycles of local discussions. In each cycle, random groups of agents are formed with size $r$ with  $r=1, 2, ...L$. Within each group all agents eventually update their own opinion to adopt the opinion which got the local majority. All agents are floaters. In case of a tie in an even size group a local majority rule does not operate and agents keep on their opinion unchanged. One cycle of local update thus leads to new proportions $p_{t+1}$ and $(1-p_{t+1})$ with

\begin{equation}
p_{t+1}= \sum_{m=\frac{r+1}{2}}^{r}  {r \choose m} p_{t}^m  (1-p_{t})^{r-m} \ .
\label{pr-odd} 
\end{equation}
for odd sizes, and for even sizes
\begin{equation}
p_{t+1}= \sum_{m=\frac{r}{2}+1}^{r}  {r \choose m} p_{t}^m  (1-p_{t})^{r-m} +\frac{1}{2} {r \choose r/2}p_{t}^\frac{r}{2}  (1-p_{t})^\frac{r}{2} \ ,
\label{pr-even} 
\end{equation}
where $ {r \choose m}\equiv \frac{r !}{m ! (r-m) !}$ is a binomial coefficient. In the even case, taking  $r=2l$, Eq. (\ref{pr-even}) can be shown to reduce to Eq. (\ref{pr-odd}) with $r=2l-1$.

\begin{figure}
\centering
\includegraphics[width=1\textwidth]{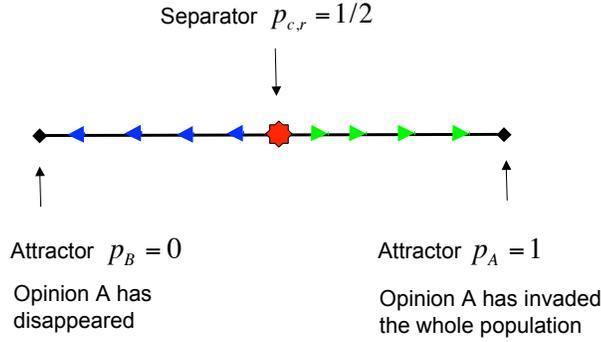}
\caption{The opinion flow for the A opinion with $p_{c,r}=1/2$.}
\label{flows}
\end{figure}

The dynamics driven from repeating these update rules is obtained solving the fixed point Equation $p_{t+1}= p_{t}$. It yields two attractors $p_{A}=1$ and $p_{B}=0$ and a  critical threshold $p_{c,r}=\frac{1}{2}$, which separates the flow opinion in direction of either $p_{A}$ or $p_{B}$. From $p_t>p_{c,r}$ there exists a number  $n$ of updates to reach equilibrium with $p_t<p_{t+1}<p_{t+2}< \dots <p_{t+n}=p_{t+n+1}=p_A=1$ where only opinion A exits. From $p_t<p_{c,r}$ there exists a number $m$ of updates to yield $p_t>p_{t+1}>p_{t+2}>\dots >p_{t+m}=p_{t+m+1}=p_B=$ where now opinion A has disappeared as seen from Figure \ref{flows}. The actual values $(n, m)$ are a decreasing function of $r$. In cases people meet in groups of different sizes a combination of Eq. (\ref{pr-odd}) is taken with the various sizes $r$ which are  involved, each weighted by the proportion of the corresponding size  \cite{mino}. 

\subsection{The mixed inflexible-floater model}

\begin{figure}
\centering
\includegraphics[width=1\textwidth]{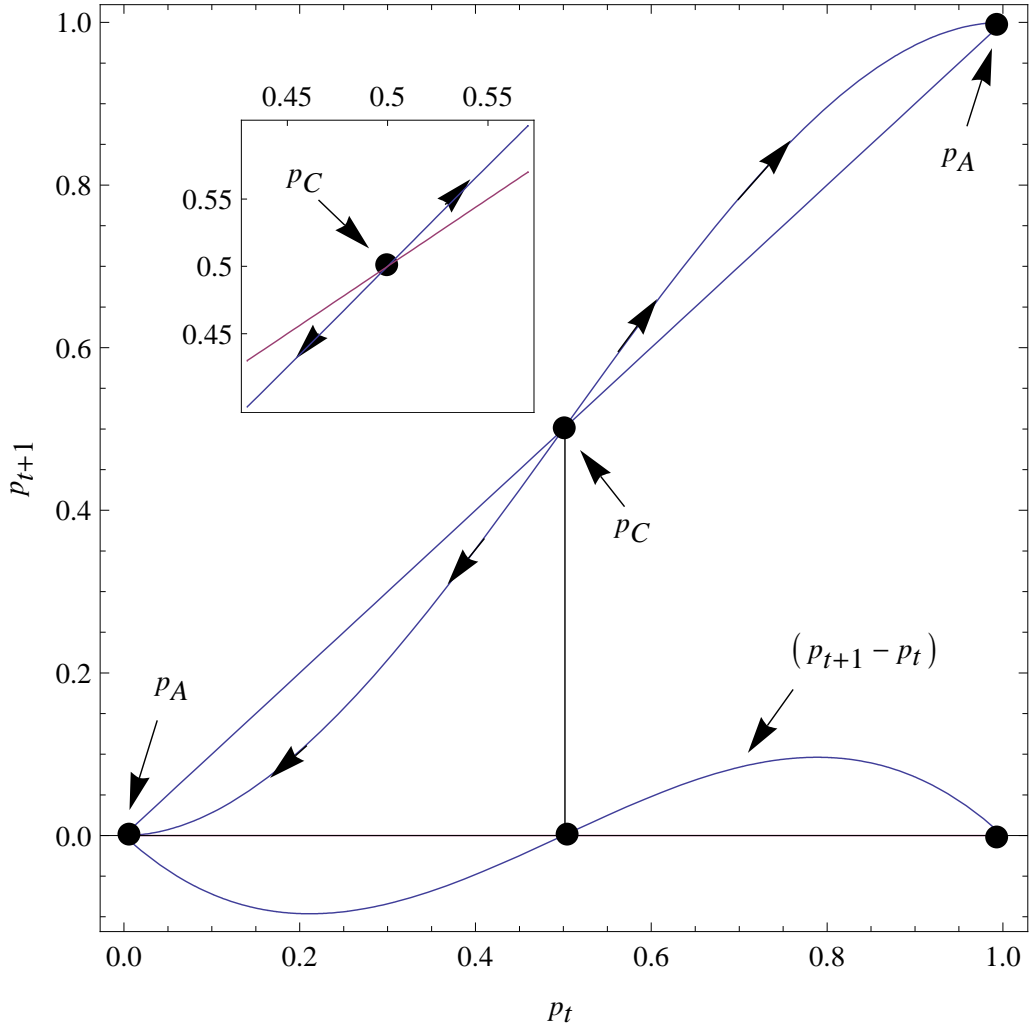}
\caption{Case $a=b=x=0$ with two attractors $p_{A}=$ and $p_B=0$ and  a separator$p_C=1/2$. The winning opinion is the one which start with a majority of initial support. 
Arrows indicate the directions of the public opinion dynamics. The inset shows the area in the vicinity of $p_C$.  The lower curve shows the evolution of the difference $(p_{t+1}-p_t)$ as a function of $p_t$.}
\label{evo0}
\end{figure}

We now investigate the effect of having inflexible agents. Denoting $a$ and $b$ the percentages of respective inflexibles for A and B the associated respective proportions of floaters are $(p_t -a)$ and $(1-p_t -b)$. We thus have $p_t\geq a$ and $1-p_t\geq b \Longleftrightarrow p_t\leq 1-b$, which combine to 
\begin{equation}
a\leq p_t \leq 1-b ,
\label{constraint-ab1}
\end{equation}
to which we add the constraints
\begin{equation}
0\leq a \leq 1  \qquad \mbox{and} \qquad 0\leq b \leq 1 \qquad \mbox{with} \qquad 0\leq a+b  \leq 1 .
\label{constraint-ab2}
\end{equation}
To keep the calculations simple we restrict the present study to groups of size 3 with Eq. (\ref{pr-odd}) writing
\begin{equation}
p_{t+1}=p_t^3+3p_t^2 (1-p_t) ,
\label{pr-3}
\end{equation}
which corresponds now to the case of zero inflexible. It  is exhibited in Figure (\ref{evo0}). The opinion which starts with a majority of initial support wins the debate. Including inflexibles with respective densities $a$ and $b$ turns Eq. (\ref{pr-3}) to
\begin{equation}
p_{t+1}=p_t^3+3p_t^2 \left [(1-p_t-b)+\frac{2}{3} b\right ]+3 (1-p_t)^2  \left [\frac{1}{3}a \right] .
\label{inflex-ab}
\end{equation}

To compare readily the difference in inflexible supports we write $a\equiv b+x$ where $x=a-b$ measures the actual difference. Eqs. (\ref{constraint-ab2}) and (\ref{inflex-ab}) become
\begin{equation}
-b\leq x \leq 1-2b ,
\label{constraint-ab3}
\end{equation}
 and
 \begin{equation}
p_{t+1}=-2p_t^3+p_t^2 (3+x)+ (-2 p_t+1) (b+x) .
\label{inflex-bx}
\end{equation}

\begin{figure}
\centering
\includegraphics[width=1\textwidth]{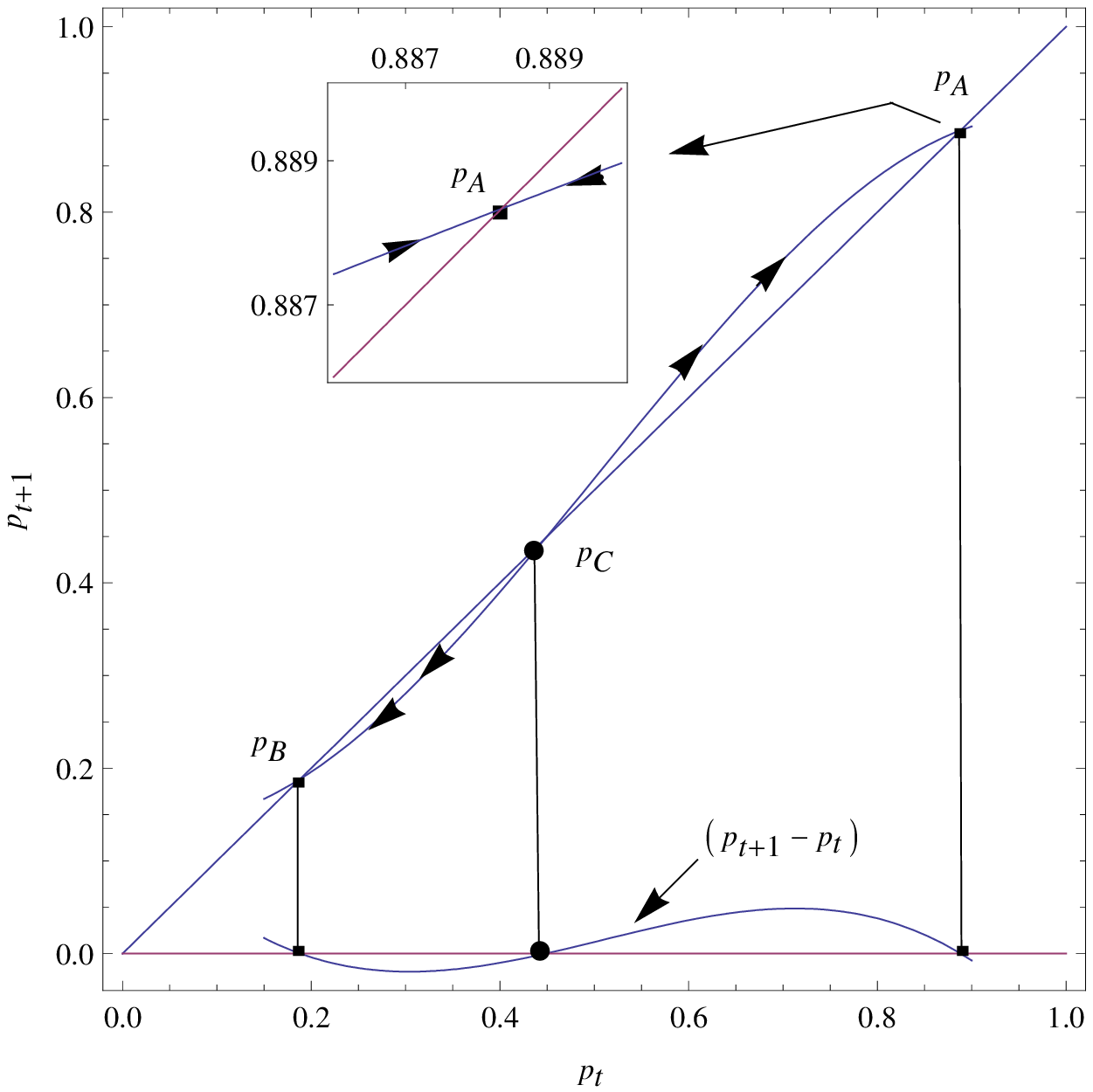}
\caption{Case $a=0.15, b=0.10, x=0.05$ yielding two asymmetric attractors $p_{B}=0.19$ (A is minority, B is majority) and $p_{A}=0.89$ (A is majority, B is minority) with a separator $p_C=0.45$ (advantage to A). Arrows indicate the directions of the public opinion dynamics. The inset shows the area in the vicinity of $p_A$. The lower curve shows the evolution of the difference $(p_{t+1}-p_t)$ as a function of $p_t$.}
\label{evo1}
\end{figure}

\begin{figure}
\centering
\includegraphics[width=1\textwidth]{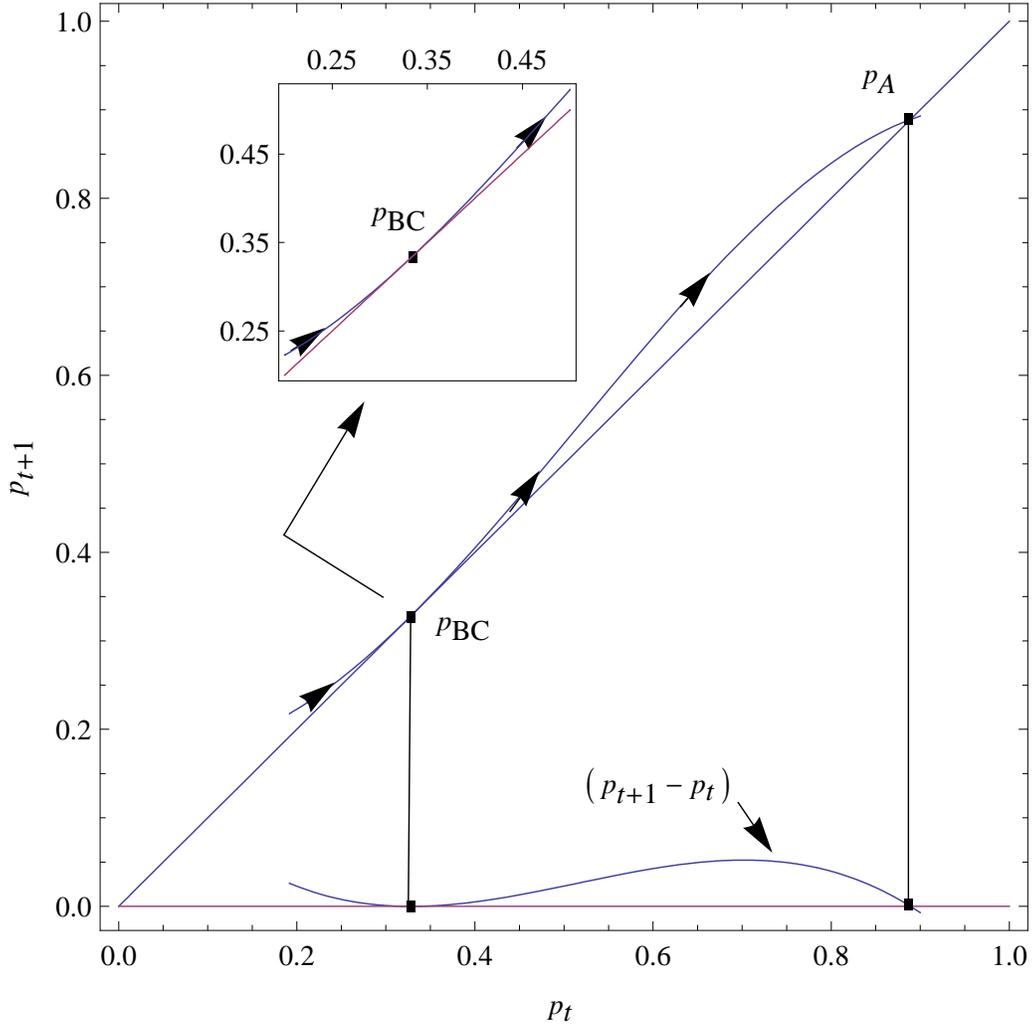}
\caption{Case $a=0.19, b=0.10, x=0.09$ where $p_{BC}=p_B=p_C=0.33$ is double and $p_A=0.89$ (A is majority, B is minority).  Arrows indicate the directions of the public opinion dynamics. The inset shows the area in the vicinity of $p_{BC}$, which is very flat making the dynamics almost stable. The lower curve shows the evolution of the difference $(p_{t+1}-p_t)$ as a function of $p_t$.}
\label{evo2}
\end{figure}

\begin{figure}
\centering
\includegraphics[width=1\textwidth]{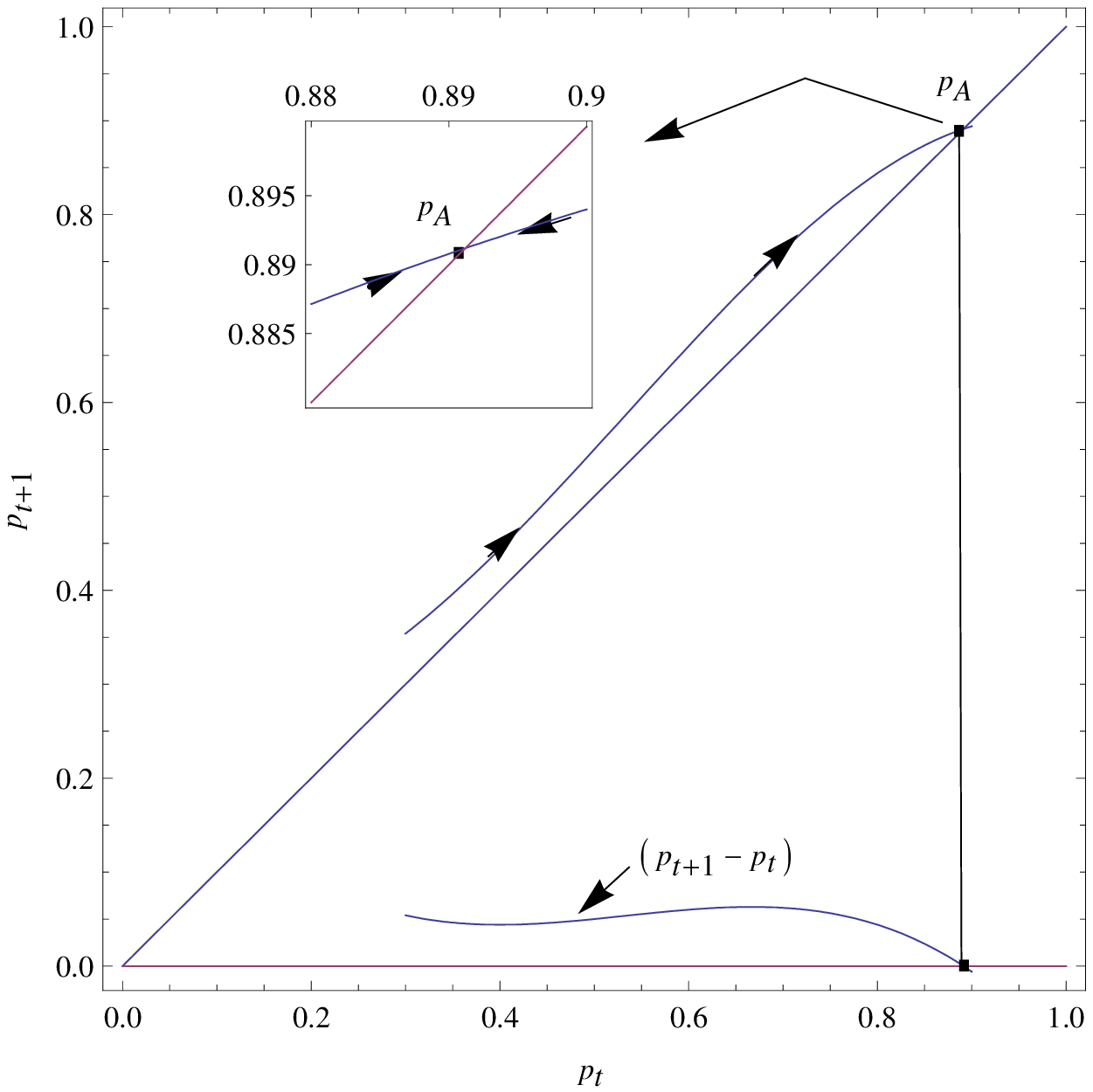}
\caption{Case $a=0.30, b=0.10, x=0.20$ with one fixed point $p_{A}=0.89$ (A is majority).  Arrows indicate the directions of the public opinion dynamics. The inset shows the area in the vicinity of $p_A$. The lower curve shows the evolution of the difference $(p_{t+1}-p_t)$ as a function of $p_t$.}
\label{evo3}
\end{figure}

The fixed points of Eq.(\ref{inflex-ab}) are real solutions of the equation $p_{t+1}=p_{t}=$, which is a cubic equation in $p_t$. It exhibits three cases depending on the sign of the associated discriminant $D$, which is a function of $b$ and $x$. Three distinct real solutions are found for $D<0$, two distinct real solutions of which one is double at $D=0$, and one single real solution when $D>0$ \cite{inflexible}. The associated flow dynamics are shown in Figures (\ref{evo1}, \ref{evo2}, \ref{evo3}).

%%%%%%%%%%%%%%%%%%

%%%%%%%%%%%%%%%%%%

When three fixed points are obtained, dynamics consistency implies that the one in between must be a separator while the two others are attractors. One case with $a=0.15, b=0.10, x=0.05$ is shown in Figure (\ref{evo1}).  The two asymmetric attractors are located at $p_{B}=0.19$ and  $p_{A}=0.89$. The first yields a B majority which coexists with a A minority while it is the opposite for the second one with a A majority with a B minority. The separator $p_C=0.45$ gives a substantial advantage to A for the democratic process of public forming since A must start with an initial support greater than $45\%$ to be sure to win the debate. The lower curve shows the evolution of the difference $(p_{t+1}-p_t)$ as a function of $p_t$. It is zero at the fixed points, negative 
when the opinion flows towards $p_{B}$ and positive when it is towards $p_{A}$.

It is worth to underline that a positive value of $x$, i.e., at the advantage of A, shifts simultaneously the separator below fifty percent and $p_B$ to a higher value since the incompressible A minority increases. The reverse holds true for negative $x$, i.e., the separator gets larger than fifty percent and $p_A$ decreases. However these shifts are not linear functions of $x$. Given $b$ there exists a critical value $x_{cA}$ at which $p_C$ and $p_B$ coalesce giving two real fixed points, one $p_{BC}=p_B=p_C$ being double. One case is exhibited in Figure (\ref{evo2}) with $a=0.19, b=0.10, x=0.09$ where $p_{BC}=0.33$ and $p_A=0.89$. By symmetry, there also exists  a critical value $x_{cB}$ at which $p_C$ and $p_A$ coalesce giving two real fixed points, one $p_{AC}=p_A=p_C$ being a double one.

In the vicinity of $p_{BC}$ the variation of $p_t$ is very flat as seen in the Figure (\ref{evo2}). It yields an extremely slow dynamics. It is a misleading geometry from which one could conclude wrongly that the opinion is stable with B as a majority. It means that while several consecutive polls would conclude an attractor has been reached, later on, all of a sudden the A opinion will start to increase quickly to become a majority as a surprise to everyone. 

An infinitesimal increase of $x$ from $x_{cA}$ erases $p_{BC}$ leaving only one unique attractor $p_{A}$. Then, any initial condition leads to the victory of A with an opinion flow leading towards $p_{A}$. Figure (\ref{evo3}) shows the case $a=0.30, b=0.10, x=0.20$. Initial conditions are irrelevant with the A opinion reaching always the majority with $p_{A}=0.89$.

When $x > x_{cA}$ or $x < x_{cB}$ the dynamics outcome is certain yielding a large majority to the opinion which has the surplus of inflexibles. It invades the majority of the population, A for the first case and B for the second one.

%%%%%%%%%%%%%%%%
\section{The instrumental key to win the debate}

From above three cases, different strategies can be elaborated to win a public debate. However, to determine the parameters for which the debate is going to be held is of a central importance. But, beforehand we can already single out two major strategies, either to act to get the largest initial support among the population or instead to focus on ``buiding" inflexibles on one's own side with eventually the de-making of other side's inflexibles. Both strategies not being exclusive. Nevertheless, it sounds reasonable to assume that to make or de-make an inflexible requires much more effort and investment than to convince a floater. It should be stressed that working on inflexible proportions modify the topology of the flow diagram while dealing with floater assumes a given topology.

\begin{figure}
\centering
\includegraphics[width=1\textwidth]{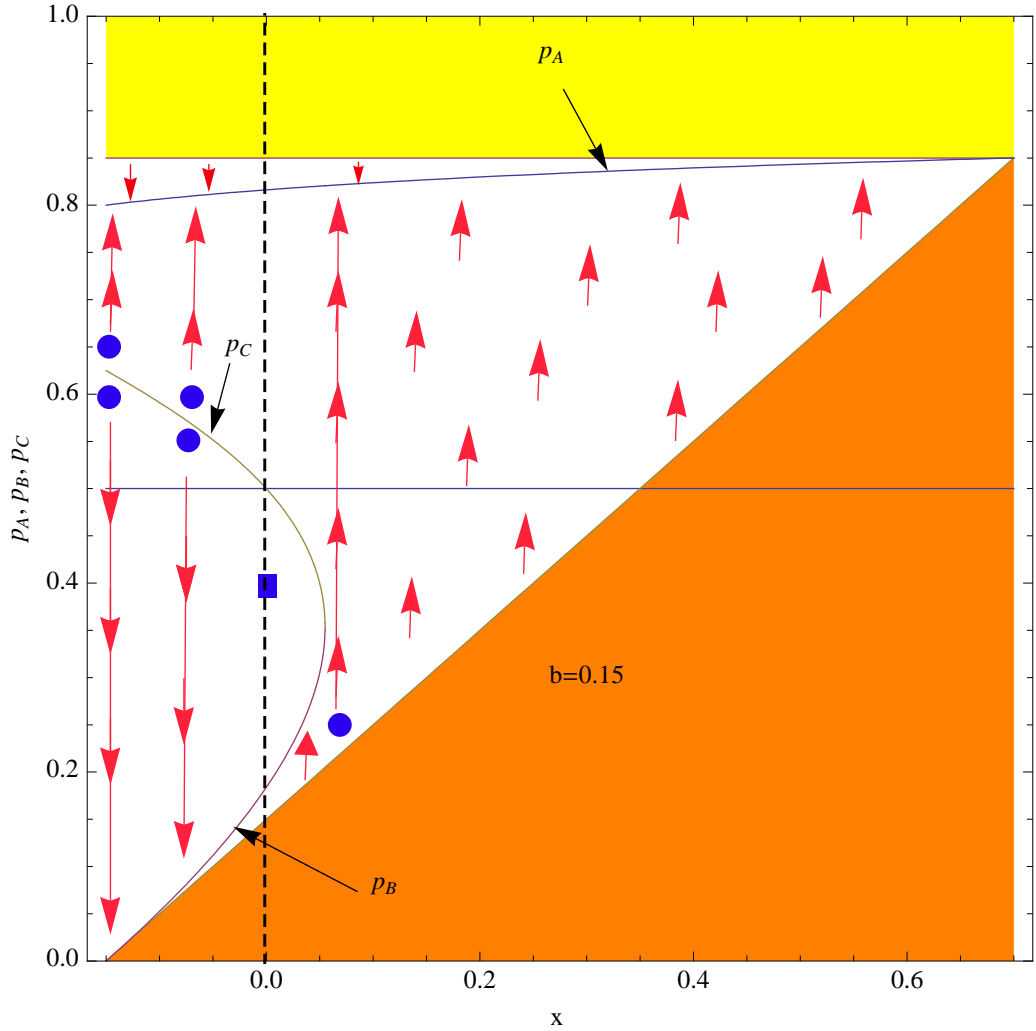}
\caption{Attractors and separator $p_{A}, p_{B}, p_{C}$ as a function of $x$ for a fixed value $b=0.15$. Filled areas represent the $p$ lower ($b+x=0.15+x$) and upper ($1-b=0.85$) values produced by both sides inflexibles. For  $x\leq 0.055$ a separator determines the faith of the public debate while in the range $x>0.055$, A is certain to reach the confortable majority of more than $80\%$, even for initial supports as low as $22\%$. Filled circles represent several initial conditions. Arrows indicate the directions of the public opinion dynamics. The square represents $p_t=0.40$ with  $a=0.15$. The associated strategies to reverse its otherwise loosing faith are discussed in the text.}
\label{triple1}
\end{figure}

Analyzing the various topologies and the way they are changed by the inflexible proportions it appears than the key sensitive issue is to ensure one has more inflexibles than the other side since even a small difference in proportions has a dramatic effect on the resulting topology of the opinion flow diagram. To substantiate that point we show in  Figure (\ref{triple1}) how the opinion flow topology varies as a function of $x$ for a given value $b=0.15$, where $b$ is the proportion of B inflexibles.

Two main regimes are present, the one with a separator $p_C$ for $x\leq 0.055$ and the one with a single attractor $p_A$ for  $x > 0.055$. The lower and upper values of $p$ are respectively $b+x=0.15+x$ and  $1-b=0.85$ due to a proportion of $b+x$ of A inflexibles and $b$ of B inflexibles. When $x\leq 0.055$ a separator determines the fate of the public debate, which means the initial conditions are crucial to eventually win or lost the public debate. In contrast for $x>0.055$ the intial conditions become irrelevant with A  certain to reach the comfortable majority of more than $80\%$, even if its initial support is low as $22\%$.

\subsection{Two different strategies}

From above topology of the opinion flow two different strategies can be elaborated in order to win the public debate. Standing from the opinion A view point the focus can be either on convincing more floaters prior to the beginning of the public debate, i.e., the initial conditions, or to concentrate on the building of inflexibles. Here these parameters are set and then the dynamics is turned on. But in reality, additional external changes can also be achieved during the public debate.

Consider for instance a case where at time $t$ we have a value $x=-0.15$, i.e., A has no inflexible while the B align $15\%$ of inflexibles. From Figure (\ref{triple1}) it is seen that for $p_t<0.625$ including $p_t>0.50$, the B opinion is certain to convince the whole population to adopt its view. Such a total victory, which includes the possible reversing of a large majority, being driven naturally by the democratic debate among all the agents. In order to win the debate A must succeed in getting more than $62.5\%$ individual support.

Another case corresponds to having the same proportion of inflexibles for A  as  for B with $a=b=0.15$. It means that $p_t$ is located somewhere on the dashed line in Figure (\ref{triple1}). The square indicates $p_t=0.35$. From this initial minority condition A is going to loose both the public debate and  most of its supporters. 

We now investigate the possible strategies given to A to reverse its fate and obtain a victory in the public debate. Several options are available as shown in Figure (\ref{triple1}). Circles indicate some modified initial position for A. It includes two changes, the  proportion $a$ of A inflexibles and the total support in the population. 

Indeed, as seen from Figure (\ref{triple1}), depending on the nature of support, i.e., its composition in terms of floaters and inflexibles, the associated dynamics outcome can be reversed or not. As long as the new $p_t$ is lower than $p_C$, even if larger than fifty percent, the A faith is set to loose the debate drastically. In order to win, A must reach an initial position either above $p_C$ or to be located in an area where $p_C$ does not exist, i.e., for $x>0.055 \Longleftrightarrow a>0.205$.

Above different strategies  can be apprehended as putting the efforts either in convincing holders of opinion B to shift to A, or to turn some A opinion holders into A inflexibles. As seen from Figure (\ref{triple1}), it might be more convenient to increase $a$ from fifteen percent up to $21\%$, i.e., to increase $x$ from zero to $0.06$, even if loosing total support down to  $21\%$. From that new initial point the public debate will drive $p$ towards more than eighty percent with a tremendous victory.

Last but not least, the floater strategy, which consists in gaining more support from floaters may be hazardous to implement, in particular in the vicinity of $p_B$ and $p_C$. A wrong evaluation like reaching $57\%$ percent instead of $60\%$ could turn the whole effort useless as seen from Figure (\ref{triple1}) with the opinion dynamics driving back the support toward minority values. The question of the respective cost of each strategy and its feasibility must be also addressed in a real implementation. It will depends on each specific case.

\subsection{The get more inflexibles winning strategy}

\begin{figure}
\centering
\includegraphics[width=1\textwidth]{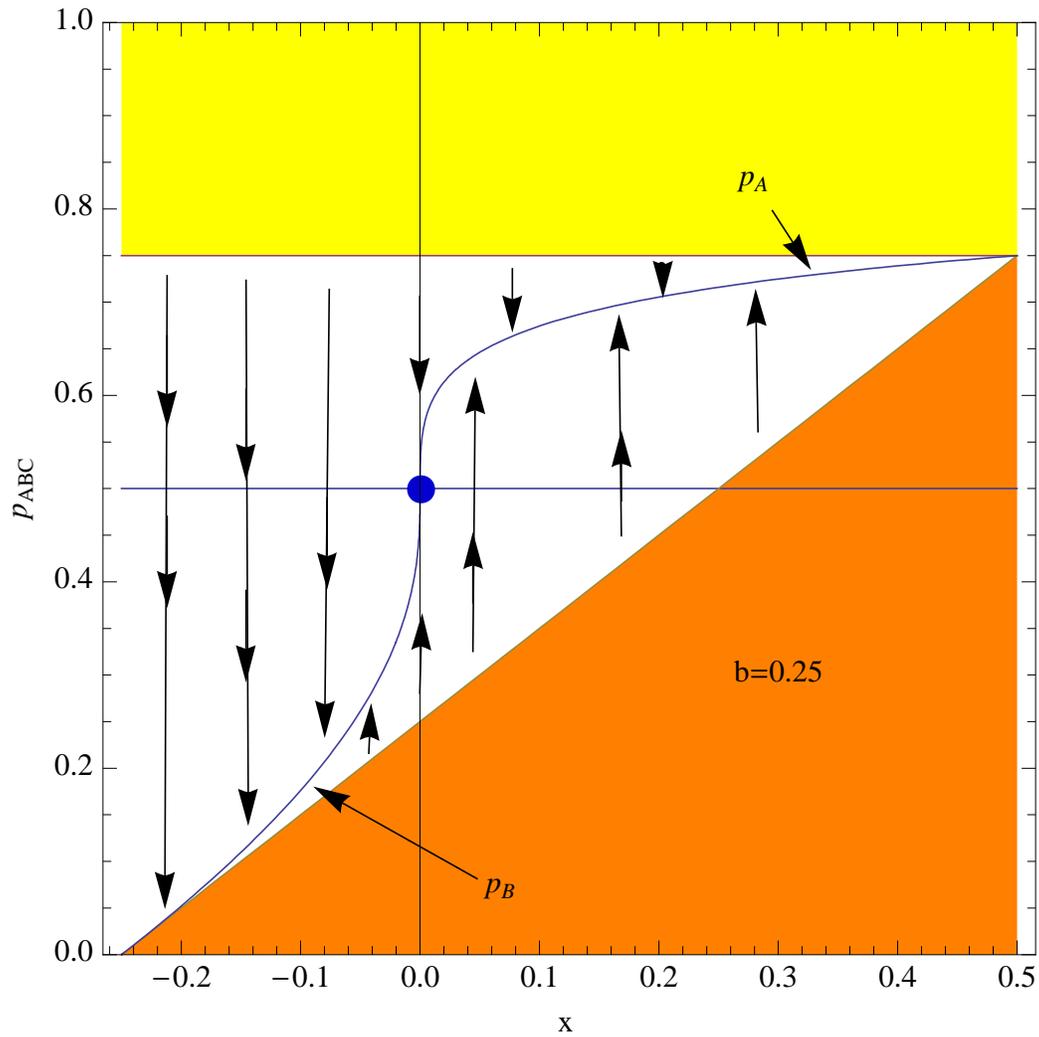}
\caption{Filled areas represent the $p_t$ lower ($p_t \geq b+x=0.25+x$) and upper ($p_t\leq  1-b=0.75$) values produced by both sides inflexibles. Arrows indicate the directions of the public opinion dynamics.}
\label{single1}
\end{figure}

Above analysis emphasizes the necessity to increase one's own side inflexibles in order to make more certain the debate outcome in one's own favor. However, with both sides acting along the same strategy, the race becomes on how to get more inflexibles than the other side. In the range $b\geq 0.25$ only one single attractor monitors the opinion dynamics flow \cite{inflexible}.  Figure (\ref{single1}) shows the opinion flow diagram at $b=0.25$ as a function of $x$. We have only $p_B$ for $-0.25 \leq x<0$ and only $p_A$ for $0<x \leq 0.50$. At $x=0$, $p_A=p_B=0.50$.

Accordingly, as long as $x<0$, B is certain to win the public debate even for very low initial supports $p_t$ of the order of a few percents. At $x=0$ the debate reaches a perfect equality. For $x>0$, A becomes the systematic winner as exhibited in Figure (\ref{single1}).

It is worth to stress that the total floater proportion decreases with increasing $x$ as exhibited by the white part of Figure (\ref{single1}). It is also essential to underline the fact that as soon as one opinion, here B, reaches a proportion of $25\%$, the only feasible option for A is to accumulate inflexibles. Failing to do so will result into a total disaster in terms of the public debate. To get a large support of floaters is meaningless within this situation. 

%%%%%%%%%%%%%%%%%

\section{Application to evolution, global warming and H1N1 pandemic influenza debates}

The above analysis along Galam model \cite{mino, hetero, inflexible} does not provide an exact explanation to the various puzzles and contradictions singled out during the discussion of the three issues of evolution theory, global warming and H1N1 pandemic influenza mentioned in Sections \ref{visit} and \ref{A}, but it sheds a new light on the different strategies used in the debates. In particular it justifies used attitudes and dismisses others opening a way to reposition some current strategies.

In the case of evolution theory it appears that inflexibles were and are  present on both sides.  Scientists refuting Intelligent Design as well as believers supporting  Intelligent Design will never change their mind since both are convinced of an indisputable truth of their respective opinion. On this basis Figure (\ref{single1}) seems to be the adequate topology to describe the associated opinion dynamics. It could explain why the debate reaches an equilibrium. The fact that the Darwin supporters got a majority would thus be the result of a larger amount of inflexibles in contrast to the Intelligent Design proponents, i.e., a positive $x$.

However the major outcome of our modeling is to confirm the rightness in overstating the validity of Darwin theory to oppose  Intelligent Design in terms of success in the public debate. Afterwards our results suggest that in case opponents to  Intelligent Design had been more circumspect about the ``status" of Darwin theory, they would have lost the public debate. It is a somehow disturbing hypothesis with an embarrassing result for an honest scientist.

For the global warming issue, inflexibles exist only on the human responsibility side, which according to our model results, makes their victory unavoidable in a public debate. A first conclusion from our results is to stress that in case alarmists had only expressed their concern about a possible man responsibility, they would have lost the debate against the skeptics as illustrated with Figure (\ref{evo0}) since in the late ninety not many people were concerned with the climate and did not consider man activities are modifying it.  Alarmists were thus well inspired on exaggerating their statements with respect to their goals. However  this result rises the ethical question of such a behavior.

In contrast, skeptics adopted a caution attitude and thus lost the debate. In case they had followed the alarmists attitude by overstating an alternative explanation to the observed global warming to challenge the claimed man responsibility, they would have certainly win the debate. But by so doing they would have exit the limit of scientific rules, the very fact they are blaming the alarmists of doing. Accordingly skeptics are trapped in a contradictory antagonism with their epistemological stand on the issue, which prevent them to make unfounded statements. They are thus bound to loose the public debate. 

From such a state of affair only a breakthrough in the understanding of the mechanisms behind global warming could reverse the current situation by dismissing alarmist claims. It would imply to provide a validated explanation to the observed climate changes. Such a scenario is a priori not expected soon due to the difficulty of the problem. On the other hand, a change on the climate trend with a continuation of the recent ongoing global cooling \cite{cooling} will reduce the public fears  and put at stake the alarmist threat. But such an hypothesis is independent of a human will. 

The only feasible option for skeptics to have a chance to reverse the opinion dynamics is to rebuke alarmists by shaking the solidity of their inflexibility.  A rather uneasy program since it is the nature of  an inflexible to be not flexible. Nevertheless the very recent publication of a large amount of private emails from climatologists, which showed some misconducts in the handling of scientific data, could result in a weakening of the alarmist inflexibility status \cite{email}. It  could drive a renewal of the public debate with an eventual change in the direction of the opinion dynamics. Indeed it is what happened in the beginning of the 2010 year when several errors were identified within the 2007 IPCC report. In particular the assertion that Himalayan glaciers are likely to disappear by 2035 is false and has produced a solid shake in the public confidence about the IPCC claims.

In the case of H1N1 pandemic influenza, the situation is more simple. Inflexibles were not numerous, but most of the existing ones were on the refusal side. In addition, most of the french population were not favorable to the full scale vaccination campaign launched by the government. While for global warming issue the initial minority inflexibles have driven a reversal of the majority to adopt their support of the IPCC view, for the H1N1 pandemic influenza, the few inflexibles , which were on the side of the initial majority did stabilize this majority against the minority, which thus remained a minority. This frame may provide an explanation to  the overwhelmed abstention of the french population to get vaccinated.

\section{Conclusion}

Putting in parallel the three issues of evolution theory, global warming and H1N1 pandemic influenza it appears that for the first  case, scientists were right in asserting Darwin theory is proved since otherwise they would have lost the debate against the  Intelligent Design proponents. A ``wrong" statement has authorized a valid outcome, i.e., to reject the ``wrongly" claimed scientific theory of Intelligent Design. 

For the second case, proponents of man responsibility in the global warming have won the public debate by making exaggerated assertions, which cannot be refuted apart using alternative exaggerated statements about a natural cause to global warming. Simultaneously, would the alarmists have defend their claims as strong presumptions instead of scientific proofs, the public opinion would have certainly reject the necessity to curb carbon dioxide emissions.

The last case of H1N1 pandemic influenza shows the key role of the resoluteness to carry out successfully the implementation of a campaign even if billions have been spent for it. The fact that the implementation is free of charge for the people and at their potential benefit is not sufficient to fulfill the campaign goal.

Our findings lead to the unfortunate and disturbing conclusion that to adopt a fair discourse is a definite lose-out strategy to promote a cause in a public debate. On this basis one could conclude that to adopt a cynical behavior is the obliged path to win a public debate against unfair and rigid opponents. However, an alternative conclusion could be to dismiss the increasing weight given to the public opinion in the process of policy making by decision makers.

It is worth to notice that the IPCC has gone through a series of set back starting before the Copenhagen summit with the so called ``Climategate" and followed by the summit failure and the discoveries of errors in the 2007 report. It has prompted a substantial shift in both the media and the public opinion which can be linked to the fact that the basis for  inflexibilty has been shaken. As a consequence the public opinion has been driven in a resuming of the debate which is going on at the moment.

Above results may enlighten the mechanisms by which the public debate about the human culpability with respect to the global warming has gained such an increasing support all over the world.  It also enlightens the ebb tide which occurred recently in the public support during the first months of 2010. Such a quick reversal looked as rather improbable up to a few month ago till the end of 2009. But that is not a formal proof of the validity of our model. It is another way to tackle the associated dynamics which on this basis deserves more investigation.

Last, but not least, sociophysics is a promising field by its specific capacity to reproduce some complex social situations within a new coherent frame. It allows the discovery of novel and counter intuitive dynamics active in the social reality. However,  it is of a crucial importance to keep in mind that we are using models to mimic part of the reality. They are only an approximation of that reality. They are not the reality. To forget the difference may lead to some misleading conclusions of what should be done to implement a policy. The limits of the approach must be always discussed before making any prediction. At this stage, the collaboration with experimentally motivated researchers from social sciences could be valuable to make predictions on selected social phenomena.

\section*{Acknoledgement}

I would like to thank Marcel Ausloos for fruitful comments on the manuscript.

%%%%%%%%%%%%%%%%%%%%%%

\end{document}